\documentclass[10pt,conference]{IEEEtran}
\usepackage{graphicx}
\usepackage{verbatim}   
\usepackage{color}      
\usepackage{subfigure}  
\usepackage{amssymb}    
\usepackage{amsmath}    
\usepackage{amsthm}     
\usepackage[dvips]{epsfig}
\usepackage{float} 
\restylefloat{table}

\usepackage{booktabs} 

\usepackage{algorithmic} 

\usepackage[numbers,sort&compress]{natbib}

\usepackage{rotfloat} 

\newtheorem{theorem}{Theorem}
\newtheorem{lemma}{Lemma}
\newtheorem{corollary}{Corollary}
\newcommand{\bolds}[1]{\mbox{\boldmath $#1$}} 

\restylefloat{figure}

\begin{document}



\title{\large A Low Rank Gaussian Process Prediction Model for Very Large Datasets}
\author{Roberto Rivera\footnote{University of Puerto Rico, Mayaguez}}
\maketitle

\pagestyle{plain} 


\begin{abstract}
	Spatial prediction requires expensive computation to invert the spatial covariance matrix it depends on and also has considerable storage needs. This work concentrates on computationally efficient algorithms for prediction using very large datasets. A recent prediction model for spatial data known as Fixed Rank Kriging is much faster than the kriging and can be easily implemented with less assumptions about the process. However, Fixed Rank Kriging requires the estimation of a matrix which must be positive definite and the original estimation procedure cannot guarantee this property. 
	
	We present a result that shows when a matrix subtraction of a given form will give a positive definite matrix. Motivated by this result, we present an iterative Fixed Rank Kriging algorithm that ensures positive definiteness of the matrix required for prediction and show that under mild conditions the algorithm numerically converges. The modified Fixed Rank Kriging procedure is implemented to predict missing chlorophyll observations for very large regions of ocean color. Predictions are compared to those made by other well known methods of spatial prediction.
\end{abstract}

\section{Introduction}
 Gaussian process prediction (GPP) , require computations of $O(n^3)$, which
represents a formidable amount of computation when $n$ is very large.
One way to overcome the computational burden has been using
regression splines or penalized regression splines when the spatial dependence is modelled
deterministically \citep{paciorek07,wood06}. In the case of the
input dependence being modeled stochastically, several models
have been suggested to reduce computational burden. Among them we
have, tapering the covariance matrix \citep{furreretal06},
approximating the field using a Gaussian Markov random field \citep{rueheld05}, and doing computations in the spectral domain
\citep{fuentes07}. Often these stochastic models assume
stationarity, an assumption that many times is justified for small
sample sizes but likely to be too restrictive for large datasets.
Moreover, published application of these methods primarily
consider scenarios with a few thousand monitoring locations. When
tens of thousands of observations or more are available, many of
the techniques become unfeasible. In this article, we aim to
modify a recent approach called Fixed Rank Kriging \cite{cressiejohannesson08}. Particularly, we give a necessary
and sufficient condition that ensures $F=C-bD$ is positive
definite. Motivated by this result, we
propose a new algorithm applied to Fixed Rank Kriging to estimate
the necessary parameters $\sigma^{2}$ and matrix $K$ while
ensuring positive definiteness of $\hat{K}$, the estimate of $K$.
In this article we denote that a matrix $F$ is positive definite
by $F \succ 0$. In sections \ref{sec:frk} to \ref{sec:frkparamest}
Fixed Rank Kriging is presented. This method makes predictions
feasible even in the case of very large datasets, without assuming
stationarity. In section \ref{sec:posdefconstraint} we introduce the new algorithm that ensures
$\hat{K} \succ 0$  and we show that this algorithm converges
numerically under general conditions. This is followed by an application using ocean color data.

\section{Fixed Rank Kriging}
\label{sec:frk} Cressie and Johannesson \cite{cressiejohannesson08} introduced a new way of obtaining spatial predictions. This method is called Fixed Rank Kriging (FRK),
and mainly addresses two drawbacks of classical GPP. One is
that GPP
cannot be implemented in the case of very large datasets since, in
general, the computations involving the inverse of the covariance
matrix, $\Sigma^{-1}_{K}$, are of the
order $O(n^{3})$. Even a few thousand observations will make the
GPP computations unfeasible. With FRK, the computations per
prediction location are significantly faster while no longer requiring a stationary assumption. This
section briefly reviews the inner workings of FRK as presented in \cite{cressiejohannesson08}, which from now on is referenced as CJ08.
The
model expressed in terms of a hidden process is,\\
\begin{eqnarray}
Y(\bolds{s}) &=& \bolds{x}(\bolds{s})'\bolds{\beta} + W(\bolds{s}) + \epsilon(\bolds{s})\nonumber\\
&=& H(\bolds{s}) + \epsilon(\bolds{s}) \label{eq:spfrk}
\end{eqnarray}
where $H(\bolds{s})=\bolds{x}(\bolds{s})'\bolds{\beta} +
W(\bolds{s})$ is the hidden process of interest. $\bolds{x}(\bolds{s})'\bolds{\beta}$
represents a random field mean and/or effects of $p$ covariates. For
observations at $n$ locations $\bolds{s}_{1}),....,\bolds{s}_{n}$,
let $\bolds{Y}= (Y(\bolds{s}_{1}),....,Y(\bolds{s}_{n}))^{'}$, and
$X$ be a $n \times p$ matrix with $k^{th}$ column
$(X_{k}(\bolds{s}_{1}),....,X_{k}(\bolds{s}_{n}))^{'}$ for
$k=1,..,p$. Assume $rank(X) = p$ with $p < n$. Furthermore, assume
$W(\bolds{s})$ and $\epsilon(\bolds{s}^{*})$ are uncorrelated for
locations $\bolds{s},\bolds{s}^{*} \in \mathcal{D} \subset
\Re^{d}$. Also assume
$Var(\epsilon(\bolds{s}_{i}))=\sigma^{2}v(\bolds{s}_{i})$ for
$i=1,...,n$ and that $W(\bolds{s})$ is a Gaussian process with
$Var(W(\bolds{s})) < \infty$ $\forall \bolds{s} \in \mathcal{D}$.
If a fixed number $r$ of basis functions are chosen such that
$\bolds{Z}(\bolds{s}) \equiv
(Z_{1}(\bolds{s}),...,Z_{r}(\bolds{s}))'$ are the basis functions
evaluated at location $\bolds{s}$, CJ08 represent the spatial
covariance of $W(\bolds{s})$ between locations
$\bolds{s}_{i},\bolds{s}_{j}$ by,
\begin{eqnarray}
C(W(\bolds{s}_{i}),W(\bolds{s}_{j})) =
\bolds{Z}(\bolds{s}_{i})^{'}K\bolds{Z}(\bolds{s}_{j})
\label{eq:frkcov}
\end{eqnarray}
using a symmetric positive definite $r \times r$ matrix $K$, a
matrix that will be estimated later on. Generally, (\ref{eq:frkcov}) is not a function of
distance between locations and is
therefore a nonstationary covariance. CJ08 use the
eigen-decomposition of $K$ to show that (\ref{eq:frkcov}) can be
interpreted as being similar to a truncated Karhunen-Loeve
expansion but with non-orthogonal
functions, $\bolds{Z}(\cdot)$.\\

Define $Z$ as the $n \times r$ matrix with $i^{th}$ row given by
$\bolds{Z}(\bolds{s}_{i})' =
(Z_{1}(\bolds{s}_{i}),...,Z_{r}(\bolds{s}_{i}))$, and let $V$ be a
$n \times n$ diagonal matrix with $v(\bolds{s}_{i})$ as the
$i^{th}$ diagonal entry ($v(\bolds{s}_{i})$ is assumed known).
Combining (\ref{eq:spfrk}) and (\ref{eq:frkcov}) gives the
covariance matrix of $\bolds{Y}$ in terms of $K$ as,
\begin{eqnarray}
Var_{K}(\bolds{Y}) \equiv \Sigma_{K} = ZKZ' + \sigma^{2}V.
\label{eq:frkcovhidden}
\end{eqnarray}
By representing the $n \times n$ covariance matrix $\Sigma_{K}$ as in
(\ref{eq:frkcovhidden}), the Sherman-Morrison-Woodbury
equation may be used \citep[page 50]{golubvanloan96},

\begin{eqnarray}
\lefteqn{\Sigma_{K}^{-1}=(\sigma^{2}V)^{-1}-}\nonumber\\
& & (\sigma^{2}V)^{-1}Z\left(K^{-1} +
Z'(\sigma^{2}V)^{-1}Z\right)^{-1}Z'(\sigma^{2}V)^{-1}\label{eq:frkinv}
\end{eqnarray}

Now calculation of
$\Sigma_{K}^{-1}$ requires only the inversion of a $r \times r$
matrix $K$ and a $n \times n$ diagonal matrix $V$. Predictions at
locations $\bolds{s}_{o}$ are now possible by using GPP with
$\Sigma_{K}^{-1}$ specified as in
(\ref{eq:frkinv}). Thus by substitution into the GPP equation 
accounting for measurement error \cite{cressie93} we get,\\
\begin{eqnarray}
\hat{H}(\bolds{s}_{o}) =
\bolds{x}(\bolds{s}_{o})^{'}\bolds{\hat{\beta}} +
\bolds{Z}(\bolds{s}_{o})'KZ'\Sigma^{-1}_{K}(\bolds{Y} -
X\bolds{\hat{\beta}})\label{eq:frkpred}
\end{eqnarray}
where
$\hat{\bolds{\beta}}=(X^{'}\Sigma_{K}^{-1}X)^{-1}X^{'}\Sigma_{K}^{-1}\bolds{Y}$,
while the mean squared
prediction error of $\hat{H}(\bolds{s}_{o})$ is,
\begin{eqnarray}
\sigma^{2}_{K}(\bolds{s}_{o}) &=&
\bolds{Z}(\bolds{s}_{o})'K\bolds{Z}(\bolds{s}_{o}) -
\bolds{Z}(\bolds{s}_{o})'KZ'\Sigma^{-1}_{K}ZK\bolds{Z}(\bolds{s}_{o})
+ \nonumber\\
&& (\bolds{x}(\bolds{s}_{o}) -
X'\Sigma^{-1}_{K}ZK\bolds{Z}(\bolds{s}_{o}))' \nonumber\\
&& (X'\Sigma^{-1}_{K}X)^{-1}(\bolds{x}(\bolds{s}_{o}) -
X'\Sigma^{-1}_{K}ZK\bolds{Z}(\bolds{s}_{o}))\label{eq:frkstd}
\end{eqnarray}
Both the covariance parameter matrix $K$ and
the scalar $\sigma^2$ need to be estimated to proceed with FRK.\\

\subsection{Basis functions} 
\label{sec:frkbasis} Smoothing
spline bases, regression spline bases, radial basis functions,
eigenvectors or wavelets are among some basis functions that could
be used with FRK. In this work, we use
the same bisquare basis functions as \cite{cressiejohannesson08}. These revolve around the concept of gridding the locations
at different resolutions and obtaining centroid
locations. Basis functions at a coarse resolution/scale, capture general
global attributes of the process. The finer the scale becomes, the
more local are the attributes captured by each basis function at
that resolution. Another benefit of the bisquare basis function is
that is has an intuitive interpretation, the closer a location
is to a centroid, then the
closer to 1 is the basis function, while the further they are, the
closer to zero is the basis function. Furthermore each basis
function has local support. For
prediction, computations of $Z'A^{-1}Z$ and $Z'\bolds{a}$ for any
$A^{-1}$ and $\bolds{a}$ are $O(nr^{2})$ for any basis function where $r$ is the total number of basis functions
\citep{cressiejohannesson08}. But when matrix $Z$ is sparse, as
is the $Z$ matrix resulting from using bisquare basis functions, then the
operations required for prediction are $O(kr^{2})$ for $k < r <<
n$.

\begin{figure}[H]
\begin{center}

\epsfxsize=3.8in \centerline{ \vbox{
\epsffile{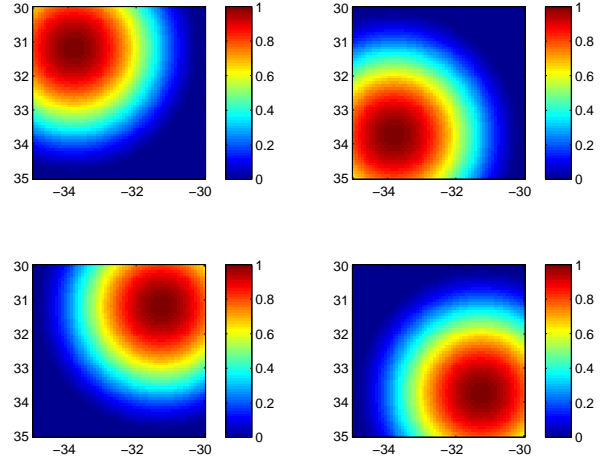}}}

\end{center}
\caption{This plot reveals the coarse scale basis functions
for a model with 2 scales of variation in a square
region of $3,600$ locations. The data is binned and the
centroid locations are obtained
according to the binning.} \label{fig:basisfunctioncoarsescale}
\end{figure}

\begin{figure}[H]
\begin{center}

\epsfxsize=3.8in \centerline{ \vbox{
\epsffile{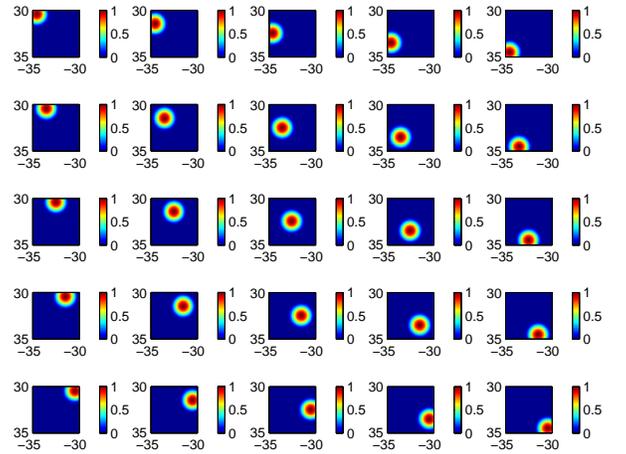}}}

\end{center}
\caption{This plot reveals the fine scale basis functions
for a model with 2 scales of variation in a square region of
$3,600$ locations. The data is binned and the centroid
locations are obtained according
to the binning.} \label{fig:basisfunctionfinescale}
\end{figure}

Suppose a FRK model is designed for a square
region of $3,600$ regularly spaced locations such that the 
covariance (\ref{eq:frkcov}) is composed of two scales of
variation, namely a coarse scale with 4 functions, and
a finer scale with 25 functions. Figure
\ref{fig:basisfunctioncoarsescale} displays the basis functions at
the coarse scale.
Similarly, Figure \ref{fig:basisfunctionfinescale} displays the basis functions at the fine scale.\\

\subsection{Parameter estimation} \label{sec:frkparamest}
Positive definiteness of the covariance matrix is
required to ensure invertibility of $\Sigma_{K}^{-1}$ and
positive mean squared prediction errors \citep{cressie93}. CJ08 bin the data for the
purpose of estimation of the covariance. This way, the estimation
and fitting of the covariance will depend on the number of bins
$M$, and not on the number of locations $n$, where $M << n$.
Assigning bin centers $\{\bolds{u}_{m}: m =1,...,M\}$, a
neighborhood $N(\bolds{u}_{m})$ is defined. The neighborhood
$N(\bolds{u}_{m})$ could be set up by some Euclidean distance as a
threshold (independent of direction, a sort of circular
neighborhood, or by distances in the horizontal and vertical
direction). Then define the
neighborhood indicator variable as,\\
\begin{eqnarray}
w_{mi} = \left\{\begin{array}{ll} 1 & \mathrm{if} \hspace{0.1cm}
\bolds{s}_{i} \in N(\bolds{u}_{m}),\
\mathrm{and}\ Y(\bolds{s}_{i}) \ \mathrm{is}\ \mathrm{not}\ \mathrm{missing},  \vspace{0.5cm}\\
       0 & \hspace{0.1cm} \rm{otherwise}\end{array}\right.\nonumber
\end{eqnarray}
Note, that a missing value at location $\bolds{s}_{i}$ would imply
a weight of zero. Furthermore, designate $\bolds{w}_{m} =
(w_{m1},...,w_{mn})^{'}$, and let $\bolds{1}_{n}$ be the $n$
dimensional column vector of ones. Without any knowledge of the
process we may use the ordinary least squares estimator
$\hat{\bolds{\beta}}$ of $\bolds{\beta}$,
to detrend the data such that, $D(\bolds{s}_{i}) =
Y(\bolds{s}_{i}) -
\bolds{x}(\bolds{s}_{i})^{'}\bolds{\hat{\beta}}$. Let $\bolds{D} =
(D(\bolds{s}_{1}),...,D(\bolds{s}_{n}))^{'}$.

 Then the average of the
residuals within the $m^{th}$ bin will be,\\
\begin{eqnarray}
\bar{D}(\bolds{u}_{m}) =
\frac{\bolds{w}_{m}^{'}\bolds{D}}{\bolds{w}_{m}^{'}\bolds{1}_{n}}
\hspace{0.5cm} \mathrm{for} \hspace{0.5cm}
m=1,...,M\nonumber
\end{eqnarray}
Let
$\bolds{\bar{D}}=(\bar{D}(\bolds{u}_{1}),...,\bar{D}(\bolds{u}_{M}))^{'}$.
An estimate of the true covariance matrix of the binned residuals,
$\Sigma_{M}=Var(\bolds{\bar{D}})$ based on the
empirical method of moments estimator has element $(m,k)$ given by,\\

\begin{eqnarray}
\widehat{\Sigma}_{M}(\bolds{u}_{m},\bolds{u}_{k}) = \left\{\begin{array}{ll} V_{D}(\bolds{u}_{k}) & \mathrm{if} \hspace{0.1cm} m=k,  \vspace{0.5cm}\\
       C_{D}(\bolds{u}_{m},\bolds{u}_{k}) & \mathrm{if} \hspace{0.1cm} m \neq
       k\end{array}\right.\label{eq:frkcovest}
\end{eqnarray}
for $m,k \in \{1,...,M\}$ where $V_{D}(\bolds{u}_{m}) =
\sum_{i=1}^{n}w_{mi}D(\bolds{s}_{i})^{2}/\bolds{w}_{m}^{'}\bolds{1}_{n}$
and $C_{D}(\bolds{u}_{m},\bolds{u}_{k}) =
\bar{D}(\bolds{u}_{m})\bar{D}(\bolds{u}_{k})$. Matrix
$\widehat{\Sigma}_{M}$ has the $(m,k)$ elements seen in equation
(\ref{eq:frkcovest}).

The intention is to approximate $\hat{\Sigma}_{M}$ by a function
of spatial dependence parameters $K$ and $\sigma^{2}$. That is, we
want a matrix $\bar{\Sigma}(K,\sigma^{2})$
with the following $\bar{\Sigma}(K,\sigma^{2})_{mk}$ entries,\\
\begin{eqnarray}
Cov(\bar{D}(\bolds{u}_{m}),\bar{D}(\bolds{u}_{k})) &=& Cov
\left(\frac{\bolds{w}_{m}^{'}\bolds{D}}{\bolds{w}_{m}^{'}\bolds{1}_{n}}
,\frac{\bolds{w}_{k}^{'}\bolds{D}}{\bolds{w}_{k}^{'}\bolds{1}_{n}}\right)\nonumber\bigskip\\
&\approx& \frac{\bolds{w}_{m}^{'}(ZKZ' +
\sigma^{2}V)\bolds{w}_{k}}
{(\bolds{w}_{m}^{'}\bolds{1}_{n})(\bolds{w}_{k}^{'}\bolds{1}_{n})}\label{eq:approxneedsadj}\bigskip\\
&\triangleq& \bar{Z}^{'}_{m}K\bar{Z}_{k} +
\sigma^{2}\bar{V}_{mk}^{*}\nonumber
\end{eqnarray}
such that if $\bar{Z} = (\bar{Z}^{'}_{1},....,
\bar{Z}^{'}_{M})^{'}$ is the $M \times r$ matrix of binned basis
functions, and $\bar{V} =
diag(\bar{V}_{11}^{*},...,\bar{V}_{MM}^{*})$, then
$\bar{\Sigma}(K,\sigma^{2}) = \bar{Z}K\bar{Z}^{'} +
\sigma^{2}\bar{V}$\footnote{Instead of deriving $\bar{V}_{mk}^{*}$ strictly from (\ref{eq:approxneedsadj}), we use $\bar{V}_{mk}=\bolds{w}_{m}^{'}V\bolds{w}_{k}/(\bolds{w}_{m}^{'}\bolds{1}_{n})$ to get a more appropriate estimator of $\sigma^{2}$. Although the $\sigma^{2}$ estimator has been rescaled,
 the resulting $\hat{K}$ remains invariant to the
 rescaling. As a consequence, FRK predictions and their standard errors are unaffected.}.  In order to approximate
$\hat{\Sigma}_{M}$ by $\bar{\Sigma}(K,\sigma^{2})$, CJ08 quantify
the 'best' approximation in terms of the Frobenius norm. Specifically for FRK,\\
\begin{eqnarray}
\|\widehat{\Sigma}_{M} - \bar{\Sigma}_{M}(K,\sigma^{2})\|_{F}^{2}
= \|\widehat{\Sigma}_{M} - \bar{Z}K\bar{Z}' -
\sigma^{2}\bar{V}\|_{F}^{2}\label{eq:frobnorm}
\end{eqnarray}
is minimized to obtain an estimate of $K$ and $\sigma^{2}$. To
ease matrix computations, the QR-decomposition of $\bar{Z}$ is
performed. That is, $\bar{Z}=QR$ where $Q$ is a column-wise
orthonormal matrix of dimension $M \times r$, and $R$ is a
nonsingular $r \times r$ upper triangular matrix. Then the $K$
that
minimizes (\ref{eq:frobnorm}) is given by,\\
\begin{eqnarray}
\hat{K}(\sigma^{2}) = R^{-1}Q'\left(\widehat{\Sigma}_{M} -
\sigma^{2}\bar{V}\right)Q(R^{-1})'\label{eq:Khat}
\end{eqnarray}
Therefore $\hat{K}(\sigma^{2})$ depends on $\sigma^2$.
Substituting (\ref{eq:Khat}) into $\bar{\Sigma}_{M}(K,\sigma^{2})$
alters (\ref{eq:frobnorm}) such that $\hat{\sigma}^2$, is the
estimator of $\hat{\sigma}^2$ that minimizes,
\begin{eqnarray}
\sum_{m,k} \left((\widehat{\Sigma}_{M} -
QQ'\widehat{\Sigma}_{M}QQ')_{mk} - \sigma^{2}(\bar{V} -
QQ'\bar{V}QQ')_{mk}\right)^{2}\label{eq:frkfrobrescale}
\end{eqnarray}
Note that remarkably, the minimization problem
for $\sigma^{2}$ is clearly of the form of a simple least squares
problem with slope $\sigma^{2}$ and no intercept and is therefore
easy to obtain (of course $\hat{\sigma}^{2}$ must be
constrained to be positive since $\sigma^{2}$ is a variance
parameter)\footnote{Although \cite[page 216]{cressiejohannesson08} state that $r \le M < n$, for the
estimation of the parameters we must have $r \neq M$. If the
number of bins $M$ were equal to the number of basis functions $r$
this would imply that also $QQ'=I$ which would make the
minimization problem (\ref{eq:frobnorm}) impossible to solve.
Therefore for FRK, it is required that $r < M <n$ ($r<M$ to ensure
column-wise orthogonality of $Q$).} \\

 When
$\sigma^{2}=0$, $\hat{K}$ is positive definite while if
$\sigma^{2} > 0$ the matrix is positive definite conditional on
the value of $\sigma^{2}$. There is no guarantee that the estimate
$\hat{\sigma}^{2}$ will result in a positive definite matrix.
In fact, Shi and Cressie (2007, page 671) state that sometimes
$\hat{\sigma}^{2}$ needs to be adjusted so that $\hat{K}$ is
positive definite but do not give further details. In this work, an algorithm is proposed to use the positive
definiteness of $\hat{K}$ as a constraint.

\begin{figure}[H]
\begin{center}

\epsfxsize=3.8in \centerline{ \vbox{
\epsffile{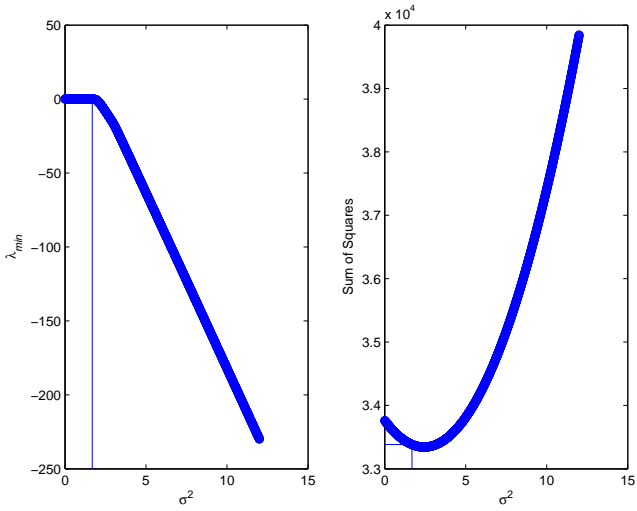}}}

\end{center}
\caption{Left panel displays the minimum eigenvalue
$\lambda_{min}$ of $\hat{K}(\hat{\sigma}^{2})$ versus different
values of $\hat{\sigma}^{2}$ plugged into (\ref{eq:Khat}).
The right panel shows the sum of squares (\ref{eq:frkfrobrescale})
for each value of $\hat{\sigma}^{2}$. Straight lines indicate $\sigma^{2}=1.6856$.}
\label{fig:simsigmavseigenminandsse}
\end{figure}

We simulate a Gaussian mean $0$ stationary
Gaussian
process with an exponential covariance function, a partial sill of 5.5, range of 1. The nugget was chosen to be 1.375, so that the signal to noise ration was 4. A sequence of
1,500 potential values of $\sigma^{2}$ were used to then calculate
$\hat{K}(\sigma^{2})$, its smallest eigenvalue $\lambda_{min}$,
and the sum of squares produced in (\ref{eq:frkfrobrescale}) by
the $\sigma^{2}$ value. $\sigma^{2}=2.4307$ is the
result of minimizing (\ref{eq:frkfrobrescale}) when only constrained so that the
estimator is positive; whereas $\sigma^{2}=1.6856$ gives the smallest sum of squares such that $\lambda_{min}>0$. Figure
\ref{fig:simsigmavseigenminandsse} shows how $\lambda_{min}$ and
the sum of squares (\ref{eq:frkfrobrescale}) related to
$\sigma^{2}$. The issue at hand is clearly seen, the minimization
of the sum of squares to find $\hat{\sigma}^{2}$ requires the
constraint that $\hat{K}$ is positive definite. Hence the solution
is a subspace of the unconstrained sum of squares problem for
$\hat{\sigma}^{2}$. Yet, positive definiteness is a quadratic
condition and appears tricky to implement as a constraint.
However, it turns out that the positive definiteness of $\hat{K}$
can be applied as a upper bound constraint on $\hat{\sigma}^{2}$.

\section{A new algorithm to estimate the FRK parameters}
\label{sec:posdefconstraint} 
Clearly from (\ref{eq:Khat}), if $\hat{\sigma}^{2}$ is too
large, then $\hat{K}$ will have negative diagonal entries. A
symmetric matrix with negative entries in the diagonal cannot be
positive definite \citep[page 141]{golubvanloan96}. We wish to
impose positive definiteness as a constraint when minimizing
(\ref{eq:frkfrobrescale}). The following lemma leads us in the right direction,\\
\begin{lemma} \label{lem:posdefdiff}
Define $F = C - bD$ where all matrices $F,C, D$ are $r \times r$
real matrices, $C \succ 0$, and $D \succ 0$, and $C$ and $D$ are
symmetric. Furthermore, assume $F$ has
distinct eigenvalues and that $b$ is any constant such that $b > 0$. Then,\\
\begin{eqnarray}
F \succ 0 \Leftrightarrow b <
\frac{\bolds{e}_{1}'C\bolds{e}_{1}}{\bolds{e}_{1}'D\bolds{e}_{1}}\nonumber
\end{eqnarray}
where $\bolds{e}_{1}$ is the eigenvector associated with the
minimum eigenvalue of $F$, $\lambda_{1}$.
\end{lemma}

The interpretation of the lemma is that, a subtraction of two
positive definite symmetric matrices of a certain form gives us a
matrix $F$ that is positive definite under certain conditions,
among them that the scalar $b$
is not too big. The following Corollary shows how $\hat{K}$ is a special case of Lemma \ref{lem:posdefdiff}.\\
\begin{corollary}\label{cor:firstcor}
Assume $\hat{K}$ in (\ref{eq:Khat}) has distinct
eigenvalues, $\lambda_{1} < .... < \lambda_{r}$. Then
$\hat{K}$ is positive definite if and only if,\\
\begin{eqnarray}
\sigma^{2} <
\frac{\bolds{e}_{1}'R^{-1}Q'\widehat{\Sigma}_{M}Q(R^{-1})'\bolds{e}_{1}}
{\bolds{e}_{1}'R^{-1}Q'\bar{V}Q(R^{-1})'\bolds{e}_{1}}\label{eq:posdefconst}
\end{eqnarray}
where $\bolds{e}_{1}$ is the $r \times 1$ normalized eigenvector
corresponding to the smallest eigenvalue $\lambda_{1}$ of
$\hat{K}$.
\end{corollary}

 Result (\ref{eq:posdefconst}) motivates the use of the
positive definiteness requirement of $\hat{K}$ as a linear
constraint on $\hat{\sigma}^{2}$.  The following algorithm
iteratively
estimates $\sigma^{2}$ and $K$.\\

\textbf{FRK parameter estimation algorithm}\\
\begin{enumerate}
\item  Calculate $Q$, $R$, $\bar{V}$, and $\hat{\Sigma}_{M}$ as described in section \ref{sec:frkparamest}.\\
\item Estimate $\sigma^{2}$ by minimizing equation
(\ref{eq:frkfrobrescale}) only subject to a constraint that
$\hat{\sigma}^{2} >0$. Start at zero an index of the iteration,
$g=0,1,..$.
Set $\hat{\sigma}^{2}_{g}$ as the result from (\ref{eq:frkfrobrescale}).\\
\item Calculate $\hat{K}_{g} \equiv \hat{K}(\hat{\sigma}_{g})$ using (\ref{eq:Khat}). \label{st:g}\\
\item Check if $\hat{K}_{g} \succ 0$. This is so if
$\lambda_{min,g} > 0$. If it is, we stop here. If it is not,
calculate an upper
bound according to (\ref{eq:posdefconst}) for $\hat{\sigma}_{g}^{2}$. Let the upper bound be $\hat{\sigma}_{u,g}^{2}$. \label{st:ubc}\\
\item Minimize (\ref{eq:frkfrobrescale}) over
$\hat{\sigma}^{2}_{g}$ but now subject to both, the greater than
zero constraint and to the upper bound
$\hat{\sigma}_{u,g}^{2}$ constraint.\label{st:ub}\\
\item Repeat steps \ref{st:g}-\ref{st:ub} above until $\hat{K}_{g}
\succ 0$. 
\end{enumerate}
This algorithm is a special case of the cutting plane algorithm
developed by Tuy \cite{tuy64,horsttuy90}. What remains is to show that as $g^{th}$ increases,
$\lambda_{min,g}$ of $\hat{K}_{g}$ increases while the upper bound
provided by (\ref{eq:posdefconst}) decreases.

Now we are ready to show the numerical convergence of the proposed FRK algorithm by stating the following theorem,\\

\begin{theorem}
\label{theorem:iterationalg} If $\lambda_{min,g}$ is the minimum
eigenvalue of $\hat{K}_{g}$ at iteration $g$, $\hat{K}_{g}$ has
distinct eigenvalues $\lambda_{min,g},...,\lambda_{max,g}, \forall
g$ and $\sigma_{u,g}^{2}$ is the upper bound found in Step
\ref{st:ubc} of the FRK
 parameter estimation algorithm at iteration $g$, then $\lambda_{min,g}
> \lambda_{min,g-1}$ if and only if $\hat{\sigma}_{g}^{2} <
\hat{\sigma}_{g-1}^{2}$.
\end{theorem}

\begin{figure}[H]
\begin{center}

\epsfxsize=3.8in \centerline{ \vbox{
\epsffile{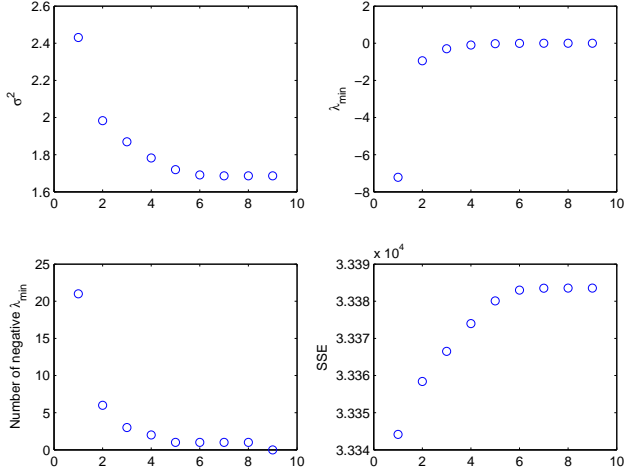}}}

\end{center}
\caption{Upper left panel displays $\hat{\sigma}^{2}_{g}$ per
iteration $g$ for simuation process $H_{1}(\bolds{s})$. Upper
right panel shows $\lambda_{min,g}$. The lower panel plots present
the number of negative eigenvalues of $\hat{K}_{g}$ and the sum of
squares (\ref{eq:frkfrobrescale}) per iteration $g$ respectively.}
\label{fig:intervalconstrainedlsqplotsforsim}
\end{figure}

Returning to the mean $0$ stationary
Gaussian
process with an exponential covariance function adopted earlier, implementing our FRK parameter estimation
algorithm using $M=100$ results in
$\hat{\sigma}^2 = 1.6861$ after 8 iterations. Figure
\ref{fig:intervalconstrainedlsqplotsforsim} demonstrates how
$\lambda_{min,g}$ and the sum of squares from
(\ref{eq:frkfrobrescale}) increase per iteration $g$, while the
number of negative eigenvalues and $\hat{\sigma}_{g}^{2}$ decrease
with every iteration.

 Kang and Cressie \cite{kangcressieshi10} propose an empirical way to ensure that $\hat{K}$ is positive definite by
 updating the eigenvalues of $\widehat{\Sigma}_{M}$. To maintain the variability features of the original
 estimate $\widehat{\Sigma}_{M}$ they must choose specific values of a parameter in their algorithm.
 In our algorithm we do not change the estimate of spatial covariance and instead iterate among the estimates of
 $\hat{\sigma}^{2}$ and $\hat{K}$. Now that we can ensure that $\hat{K}_{g} \succ 0$ the
following steps are taken to attain
the FRK prediction of a random field of interest:\\
\begin{enumerate}
\item Calculate the ordinary least square estimate
$\bolds{\hat{\beta}} =(X^{'}X)^{-1}X^{'}\bolds{Y}$ of
$\bolds{\beta}$. 
\item Obtain residuals $\bolds{D} = \bolds{Y} -
X\bolds{\hat{\beta}}$ according to the ordinary least squares
estimate of $\bolds{\beta}$.\\

\item Set up the matrix of basis functions $Z$. 

\item Next the stage is set to estimate the covariance matrix
$\Sigma_{K}$ as in (\ref{eq:frkcovhidden}). 

\item Calculate the
method of
moments estimate $\widehat{\Sigma}_{M}$ of $\Sigma_{M}$ using $\bar{D}(\bolds{u}_{m})$

\item Obtain $\bar{Z}$ and $\bar{V}$. Then perform the
QR-decomposition of $\bar{Z}$.

\item Estimate $\sigma^{2}$ and $\hat{K}$ according to the
algorithm proposed in this section and shown to numerically
converge in Theorem \ref{theorem:iterationalg}.

\item Predict the random field $H(\bolds{s}_{o})$ using
(\ref{eq:frkpred}) and its standard error with the square root of
(\ref{eq:frkstd}).
\end{enumerate}

Several things are worth mentioning. First, it is implicitly
assumed that the binning has been performed in such a way that
missing values are not present in the binned version of the data.
Otherwise weighting of the covariance estimates would not be
possible (an empty bin results in $V_{D}(\bolds{u}_{m})=\infty$
for that bin as seen from (\ref{eq:frkcovest})). More importantly
in the presence of missing values the positive definiteness of the
matrix could no longer be guaranteed. A remedy could be to remove
empty bins from (\ref{eq:frkcovest}) or impute them by some rule.
$x,y$ coordinate based averages or local averaging are just some
possibilities of imputing missing values. Cressie \cite{cressie93}
discusses median polish, another tool that could be used for
imputation of missing bins. Little and Rubin \cite{littlerubin02} provide a much
broader emphasis on ways to
deal with missing values.\\
Another noteworthy fact is that storage of the $n \times n$
$\Sigma^{-1}_{K}$ is often required in classical geostatistics, a
task of the order of $O(n^{2})$. In addition to savings on
computational time, FRK provides substantial reduction in storage.

Another point
worth mentioning is that sometimes some bins may be considerably
more variable than others. To account for differences in
variability among the bins, the influence of potential outliers,
and for bins that have more data, the parameter estimation can be
weighted. This is analogous to the use of weighted least squares
instead of ordinary least squares in variogram fitting for the
same purpose. CJ08 also took this into consideration
and based on \citep[see][page 95]{cressie93} and references therein suggest,\\
\begin{eqnarray}
\hat{a}_{m} \approx
2^{-1/2}(\bolds{w}'_{m}\bolds{1}_{n})^{1/2}/V_{D}(\bolds{u}_{m})\nonumber
\end{eqnarray}
as a weight. This translates into a rescaling of the minimization
problem (\ref{eq:frobnorm}). If $\bar{A} =
diag(\hat{a}_{1},....,\hat{a}_{M})$, then the norm is equivalent
to $\|\bar{A}^{1/2}\hat{\Sigma}_{M}\bar{A}^{1/2} -
\bar{A}^{1/2}\bar{\Sigma}_{M}(K,\sigma^{2})\bar{A}^{1/2}\|_{F}^{2}$ and,
\begin{eqnarray}
\hat{K}_{a} = R^{-1}_{a}Q^{'}_{a}\left(\widehat{\Sigma}_{M} -
\sigma^{2}\bar{V}\right)Q_{a}(R_{a}^{-1})'\label{eq:Khatwgted}
\end{eqnarray}
where we used $\bar{A}^{1/2}\bar{Z}=Q_{a}R_{a}$.
Therefore virtually no computational cost is added by using the weights. Furthermore, the previous results in this article still hold.
\begin{corollary}
Assume $\hat{K}_{a}$ in (\ref{eq:Khatwgted}) has distinct
eigenvalues, $\lambda_{1} < .... < \lambda_{r}$. Then
$\hat{K}_{a}$ is positive definite if and only if,\\
\begin{eqnarray}
\sigma^{2}_{a} <
\frac{\bolds{e}_{1}'R^{-1}_{a}Q^{'}_{a}\bar{A}^{1/2}\widehat{\Sigma}_{M}\bar{A}^{1/2}Q_{a}(R_{a}^{-1})'\bolds{e}_{1}}
{\bolds{e}_{1}'R_{a}^{-1}Q_{a}^{'}\bar{A}^{1/2}\bar{V}\bar{A}^{1/2}Q_{a}(R_{a}^{-1})'\bolds{e}_{1}}\nonumber
\end{eqnarray}
where $\bolds{e}_{1}$ is the $r \times 1$ normalized eigenvector
corresponding to the smallest eigenvalue $\lambda_{1}$ of
$\hat{K}_{a}$.
\end{corollary}
The proof is almost identical to Lemma \ref{lem:posdefdiff} and Corollary \ref{cor:firstcor}.
Theorem \ref{theorem:iterationalg} also still holds.\\

Fixed Rank Kriging is a method intended for very large datasets.
Comparisons with GPP are made difficult
for two reasons. First, if the dataset is too large, classical
GPP is not feasible. On the other hand if a dataset
is small enough for GPP, it might be too small to estimate the
spatial covariance properly using (\ref{eq:frkcovest}). For
example if $n=3,600$ one should probably have at least 30 pixels
per bin, which doesn't leave much room for a multiresolution
basis.\\
Finally since the inverse computation is not an $O(n^3)$ anymore when FRK is used,
a Likelihood approach that would more fully take into account the
probability distribution of the data in comparison to the method
of moments approach is plausible. Yet keep in mind that in
this case one of the parameters of interest, $K$, is a matrix.
Stein \cite{stein08} fits a covariance of the form (\ref{eq:frkcovhidden})
by parameterizing $K$ and then maximizing the likelihood.\\



\section{Chlorophyll data}
\label{sec:oceanmods}
 Ocean color observations enable scientists to study
several biological and biogeochemical properties of the oceans. In
part, ocean color can measure surface phytoplankton (microscopic
ocean plants) since color in most world's oceans in the visible
light region (wavelength of 400-700nm) varies with the
concentration of chlorophyll and other components, i.e the more
phytoplankton present, the greater the concentration of plant
pigments, ergo the greener the water \citep{siegeletal05a}. Ocean color is crucial for: the study of organic matter
produced by algae and bacteria, the study of the biochemistry of
the ocean, the assessment of the role of the ocean in the carbon
cycle, and the potential global warming trend. 

Based on prior and ongoing ocean color satellite missions,
scientists can now study the spatial and temporal variability of
the biological, chemical and physical processes that regulate
ocean color around the globe. For example, \cite{doneyetal03}  and
\cite{fuentesetal00} present studies of the spatial correlation
of chlorophyll at the mesoscale (about 10-200km). Substantial
progress has been made in analyzing these ocean color satellite
data, See \cite{siegeletal02, siegeletal05a, maritorenaetal02}. But many computationally
intensive statistical challenges remain. The ocean color satellite
datasets are massive, with hundreds of thousands of observations
or more around the globe. Ocean processes are generally
non-stationary in both space and time. Furthermore, due to many
factors,
satellite data comes with large amounts of missing data.\\
In this section our goal is to compare the prediction capability
of of the modified Fixed Rank Kriging (FRK) algorithm , to other well known prediction methods,
namely: ordinary least squares, regression thin plate splines, and
universal kriging, whenever computations required for the latter
are feasible. 

\subsection{Satellite data}
Satellites allow us to gather environmental data on a truly global
scale. Currently, there are several satellites sending ocean color
data to different agencies from several countries (See
http://www.ioccg.org/sensors\_ioccg.html). NASA is in charge of
two of these satellites that carry instruments sending ocean color
data. The Sea-viewing Wide Field-of-view Sensor (SeaWiFS) is one
of them, and was deployed in 1997. The other is the Moderate
Resolution Imaging Spectroradiometer (MODIS-Aqua), an instrument
available in the Aqua satellite operating since 2002. These
sensors offer different designs, orbits, and accuracy of ocean
color data. Specifically, the sensors measure the water leaving
radiance, $L_{wN}(\lambda)$, a subsurface radiance reflected out
of the ocean through an air-sea interface. Both SeaWiFS and
MODIS-Aqua measure the water leaving radiance $L_{wN}(\lambda)$ at
several wavelengths $\lambda$. These two instruments offer
different space-time resolutions, different sampling patterns, and
different measurement uncertainties. Table
\ref{tab:seawifsvsmodis} summarizes some traits that characterize
each of the two sensors.
 \\
\begin{table}[H]
\begin{center}
\begin{tabular}{|l||c c|}\hline
Sensor & SeaWiFS & MODIS\\\hline Satellite & Orbview-2 &
Aqua\\ Data available since & 09/04/1997 & 07/04/2002\\
Time equator is crossed & 12:20pm & 1:30pm \\
Spectral bands & 8 & 36 \\ Swath width & 2806 km & 2330 km \\
Spectral coverage (nm) & 402-885 & 405-14,385\\
Resolution & 1100 m& 1000 m
\\ Tilt
 & $-20^{o}, 0^{o}, 20^{o}$ & None \\
 Orbit & Descending & Ascending \\\hline\hline
\end{tabular}\\
\caption{A side to side comparison of some properties of the two
ocean color measuring instruments SeaWiFS and MODIS. Adapted from
\cite{pottieretal06} and
http://www.ioccg.org/sensors/current.html.}
\label{tab:seawifsvsmodis}
\end{center}
\end{table}

The satellite raw data plus instrument data ('Level 1') is
calibrated and reprocessed to give geophysical values, Level 2,
derived after applying several algorithmic and atmospheric
adjustments (http://oceancolor.gsfc.nasa.gov). The Level 2 data
are spatially and temporally averaged to give Level 3 statistical
data. For example, SeaWiFS has a $9km \times 9km$ spatial
resolution at the equator for Level 3 products while MODIS has
$4.6km \times 4.6km$ spatial resolution at the equator for Level 3
products. See \cite{robinsonetal00}.

The $L_{wN}(\lambda)$ data used for our analysis is processed by a
semi-analytical model developed by Garver, Siegel and Maritorena,
GSM01 \citep{maritorenaetal02}. The algorithm expresses the water
leaving radiance in terms of CHL and the Inherent Optical
Properties (IOP): CDM and BBP. In short, the GSM01 model inverts
observations of the normalized water leaving radiance spectrum,
$L_{wN}(\lambda)$, to estimate, Chl, CDM and BBP \citep{siegeletal05b}.

\begin{figure}[H]
\begin{center}

\epsfxsize=3.8in \centerline{ \vbox{
\epsffile{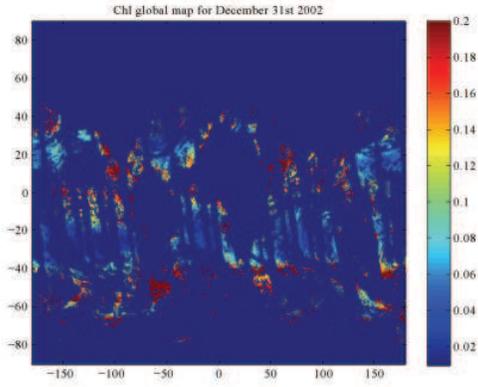}}}

\end{center}
\caption{Global chlorophyll values for December 31st 2002 using
the Aqua Sensor.} \label{fig:globalmapchldec312002aqua}
\end{figure}

Data from two NASA satellites are available. Specifically for
SeaWiFS, the time period of September 04, 1997 through July 04,
2007 is available. Aqua results from July 04, 2002 through July
04, 2007 are also available. Missing observations result from
orbital sampling, sun glint and cloud cover. Regions of the globe
often have observations for only about 30\% of the days (sometimes
less) across all years. Figure \ref{fig:globalmapchldec312002aqua}
shows the satellite measurement pattern for the Aqua sensor for
one day (December 31, 2002), and highlights the space-time missing
data problem. The orbital track of the satellite is clearly
visible (especially in the Southern Hemisphere), yet on occasion,
regions that fall along the sampling track do not have
observations. In fact, there are very few observations north of
$40^{o}$N, while more data is present south of $40^{o}$S. This
pattern in the missing values is due to a seasonal effect on the
mechanism of missing values. We focus on open ocean waters where
GSM01 output is considered to be more reliable than that coming
from coastal water
reflectance \citep{siegeletal05b}.\\
Maritorena and Siegel \cite{maritorenasiegel05} combine the $L_{wN}(\lambda)$ values
from SeaWiFS and Aqua combined to produce CHL and IOP data. This
results in better spatial/temporal coverage and sometimes lower
uncertainty surrounding the variables obtained (pixels with
multiple observations are imputed a weighted average of satellite
data according to their respective uncertainties). More recently \cite{maritorenaetal10} included MERIS data as well in the
merging procedure. We begin the analysis on the field of
chlorophyll values and the chlorophyll
dependence on spatial location.

\section{Chlorophyll for a very large region in the North and
South Pacific}
In this section, we examine the predictions obtained from four
prediction methods: ordinary least squares (OLS), an additive
model (AM), and fixed rank kriging (FRK). 
For this analysis, the observations consist of 8-day Sea-viewing
Wide Field-of-view Sensor (SeaWiFS) and Moderate Resolution
Imaging Spectroradiometer (Modis) Aqua merged data resulting from
the GSM01 algorithm. Campbell \cite{campbell95} discusses how chlorophyll
follows approximately a lognormal distribution. We follow this
result and model the chlorophyll on the natural logarithm scale
and denote
 $Y(\bolds{s}_{i}) =log(CHL_{i})$ for location (pixel) $i$.
 Hypothesis tests on
$\bolds{\beta}$ in a preliminary OLS model with
$\bolds{x(s_i)}=(1, s_{i1}, s_{i2})'$ as the $i^{th}$ row of the
$n\times 3$ matrix $X$ where $\bolds{s}_i=(s_{i1},s_{i2})$ for
$i=1,\ldots,n$ (where
$n=3600$), and with $\bolds{\beta}=(\beta_1,\beta_2,\beta_3)'$, implied a significant effect in both North-South and East-West directions.\\

Due to the size of the data, the AM is implemented using thin plate regression
splines. More precisely,
the basis functions for the thin plate regression splines model
are chosen by eigen-truncating the matrix with entries coming from
the basis functions of thin plate splines \cite{wood06}. 
The FRK procedure is implemented in Matlab using a 32-bit Linux
machine.

\subsection{Measure of predictor performance}
 $n_{t}$ observations are put aside for testing the performance of a
predictor. The remaining observations are used to fit OLS, AM,
and FRK models. Then the estimated mean squared prediction
error
can be estimated by cross validation,\\
\begin{eqnarray}
\widehat{MSPE}(\Upsilon_{m}) = \frac{1}{n_{t}}\displaystyle
\sum_{i'=1}^{n_{t}}(Y(\bolds{s}_{i'}^{*}) -
\hat{H}_{m}(\bolds{s}_{i'}^{*}))^{2}\label{eq:mspemeas}
\end{eqnarray}
calculated at test locations $\bolds{s}_{i'}^{*}$ for $n_{t}$ prediction locations, 
 and $\Upsilon_{m}$ is the $m^{th}$ modeling procedure. We also note that
(\ref{eq:mspemeas}) must be used with caution. Care is needed,
since this method does not prove that a spatial model is
correct, only that it is 'not grossly incorrect' \cite{cressie93}.

\subsection{Satellite data for regions of interest}
\begin{figure}[H]
\begin{center}

\epsfxsize=3.8in \centerline{ \vbox{
\epsffile{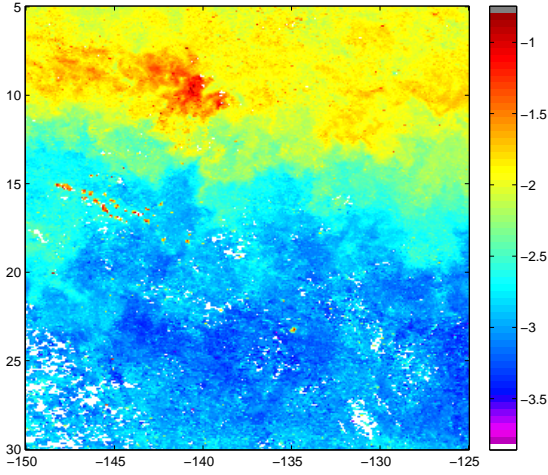}}}

\end{center}
\caption{This plot presents the log of CHL for the region between
$125^{o}$ and $150^{o}$ W and between $5^{o}$ and $30^{o}$ S.}
\label{fig:imagechlfield0530S125150W}
\end{figure}

 The regions in the
analysis are 130-155W by 5-30N in the North Pacific and 125-150W
by 5-30S in the South Pacific, both which include 90,000
locations. Kriging would require the storage of a $90000 \times
90000$ covariance matrix for each region and the inversion of this
matrix is $O(90000^3)$. In the North Pacific region the 8-day data
has 1947 missing values for the time period starting at Julian day
73 of year 2003, and in the South Pacific 2,403 missing values for
the time period starting at Julian day 177 of year 2007. These
time periods are conveniently chosen so that we can separate
different sizes of test data for the comparison of the prediction
models like we did in the previous section. Figure
\ref{fig:imagechlfield0530S125150W} displays the South Pacific
region of interest. Specks of pixels with apparent high values
relative to neighbors can be seen around $15^{o}$S, $145^{o}$W.
These are due to the coasts of the French Polynesia islands in
that region. Clearly most of the variability occurs in a
North-South direction, with an increasing trend in the northward
direction. The North Pacific region also shows most of the variability in the North-South
direction (but with an increasing trend in the southward direction).
Based on these preliminary images, we assume a linear trend with the coordinates in each direction.\\

 The AM is
implemented using thin plate regression splines. Exploratory analysis lead us to
choose 100 basis functions to fit the thin plate regression spline model.\\
Fixed Rank Kriging is implemented using bisquare basis function
with three scales of variation 16, 64, and
225 functions respectively. Parameters $K$ and $\sigma^{2}$ were
estimated using $M=900$, $V=I$ and following the new procedure
detailed in section \ref{sec:frk}.

 Missing values are omitted before the comparison of the
predictors. Then 15\% of the data are randomly removed to test
each model. The remaining observations are used to fit OLS, AM and
FRK models and then each model is used to predict the test data.
The procedure is repeated 50 times. Then the $\widehat{MSPE}$ is
obtained using (\ref{eq:mspemeas}).  The study was conducted also
using 25\% and 50\% of the observations as test data. Table
\ref{tab:modelcompnorthpacif} presents the results of the analysis
for the North Pacific. While Table \ref{tab:modelcompsouthpacif}
does so for the South Pacific. OLS gives the highest
estimates of $MSPE$. For both the North and South Pacific we find that FRK outperforms not only OLS but AM as well. The FRK
method takes about 167 seconds to calculate $\hat{K}$ and
$\hat{\sigma}^{2}$ using the procedure introduced in section
\ref{sec:posdefconstraint} and estimate the process at all 44,026
locations in the North Pacific (see Table
\ref{tab:modelcompnorthpacif}) when 50\% of the original data is
used to test the prediction method. A regression thin plate spline
implements a basis function smoothed equally in all directions.
Anisotropy in the large region could hinder any method based on
such an assumption. Yet a tensor product of regression splines \citep{wood06}, in both spatial directions
did not improve on the results of the thin plate regression
splines shown here.

\begin{table}[H]
\begin{center}
\setlength{\tabcolsep}{0.5pt} 
\begin{tabular}{|l||c|c|c|c|}\hline
\bolds{Model} & $\bolds{\widehat{MSPE}}$ (15\%)& $\bolds{\widehat{MSPE}}$ (25\%) & $\bolds{\widehat{MSPE}}$ (50\%) & CPU time\\[0.5ex]\hline
OLS & 0.0515 (6.50e-4) & 0.0517 (5.75e-4)& 0.0516 (3.63e-4) & 0.27
\\\hline AM
 & 0.0169 (3.54e-4) & 0.0169 (2.71e-4) & 0.0169 (1.93e-4) & 140.84\\\hline FRK  & 0.0099 (2.73e-4) & 0.0100 (1.88e-4)& 0.0100 (8.73e-5) & 167.68\\\hline
\end{tabular}\\
\caption{Comparison of the mean squared prediction error obtained
from OLS, AM and FRK models for the region in the North Pacific
when 15, 25 and 50\% of the data are removed and used as test
data. Each cell contains the mean $\widehat{MSPE}$ for all 50
simulated test data and the corresponding standard deviation in
parenthesis.} \label{tab:modelcompnorthpacif}
\end{center}
\end{table}

\begin{table}[H]
\begin{center}
\begin{tabular}{|l||c|c|c|}\hline
\bolds{Model} & $\bolds{\widehat{MSPE}}$ (15\%)& $\bolds{\widehat{MSPE}}$ (25\%) & $\bolds{\widehat{MSPE}}$ (50\%)\\[0.5ex]\hline
OLS & 0.0598 (8.00e-4) & 0.0597 (5.45e-4)& 0.0597
(4.21e-4)\\\hline AM
 & 0.0188 (7.33e-4) & 0.0188 (5.16e-4) & 0.0188 (3.04e-4)\\\hline FRK  & 0.0159 (6.83e-4) & 0.0159 (4.73e-4)& 0.0159 (2.63e-4)\\\hline
\end{tabular}\\
\caption{Comparison of the mean squared prediction error obtained
from OLS, AM and FRK models for the region in the South Pacific
when 15, 25 and 50\% of the data are removed and used as test
data. Each cell contains the mean $\widehat{MSPE}$ for all 50
simulated test data and the corresponding standard deviation in
parenthesis.} \label{tab:modelcompsouthpacif}
\end{center}
\end{table}

\begin{figure}[H]
\begin{center}

\epsfxsize=3.8in \centerline{ \vbox{
\epsffile{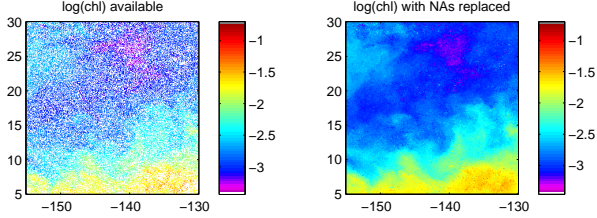}}}

\end{center}
\caption{Natural log of CHL for the region between $130^{o}$ and
$155^{o}$ W and $5^{o}$ and $30^{o}$ N for 8-day data starting on
Julian day 169 of year 2007 (left panel). Over 50\% of the
observations are missing. The observations available are used to
fit a FRK model with $M=900$ bins and $r=305$ basis functions. The
right panel shows the result of filling in the missing values
using FRK predictions.}
\label{fig:chlobswNAvschlobsandfilledNA0530N130155W}
\end{figure}

Moreover, the mean $\widehat{MSPE}$ of the prediction methods
almost do not change as more missing values are present in the
data. In the case of the OLS and AM this may be so because, as
suspected from Figure \ref{fig:imagechlfield0530S125150W}, most of
the variability occurs in the North-South direction and, if
thought of deterministically, the association between chlorophyll
and
location is almost linear. Even when half of the data is removed for cross-validation, the amount of remaining observations may be
enough that the FRK predictor performs well. Further study would be needed to confirm this.\\
FRK allows quick prediction of missing values in a massive region
without assuming that the spatial association is stationary.
Figure \ref{fig:chlobswNAvschlobsandfilledNA0530N130155W} displays
how the ocean color image would look like if missing values where
predicted using FRK and imputed into the original image (right
panel). The prediction method does not induce any unwanted
discontinuities or extreme outliers in the image. Unfortunately,
with the number of observations used in this section GPP
cannot be used. We suspect that it will underperform FRK because in this scenario the
spatial association may vary considerably across the region.

\section{Final Remarks}
FRK can be viewed as a Gaussian process random
effects model that allows prediction in the case of very large
datasets without assuming stationarity and without storing the $n
\times n$ covariance matrix. We give a result showing when the difference of two matrices (of a certain
form) is guaranteed to be a positive definite matrix. Based on
that result, we have proposed an iterative algorithm that finds
$\hat{\sigma}^{2}$ such that $\hat{K}$ is ensured to be positive
definite. We prove that the algorithm always converges
numerically. The results hold when parameter estimation is weighted according to bin variability.  
 
In this work we implement the modified FRK algorithm to ocean color data. We find that the
prediction method gives smooth predictions without assuming
stationary spatial dependence and can make these predictions
quickly, even when the data are very large. In comparison, thin
plate regression splines require more computational time while,
depending on the pattern and amount of missing data, not
performing as well as FRK. Moreover, the FRK predictions appear to hold the association between CHL and the IOPs (results not shown).\\
Many directions are possible for
future work. It would be useful to see the performance of FRK when
different basis functions and different values of $r$ are used.
Shi and Cressie \cite{shicressie07} use wavelets as basis functions. Yet
efficient implementation of wavelets requires the absence of
missing values. The authors impute the missing values to obtain
the basis function matrix. How this step affects their final
results is unclear and further study is warranted. We conjecture
that a flexible basis function can be derived with the use of
truncated tensor product basis functions. Spatio-temporal and
multivariate extensions would also be useful. Other explanatory variables known to affect ocean
color could be included in the prediction model: sea surface
temperature, wind and in situ measurements of ocean color could be
used. The effect of some of the properties ($M$, $r$, basis
function class, etc.) of the FRK model on ocean color predictions
needs further study. The FRK predictions of missing values will
provide more coverage of the ocean. The biological, chemical and
physical forcings could be studied at different spatial scales,
taking into account the predicted observations and the uncertainty
surrounding their estimation.

\appendix
\section{Proofs}
\subsection{Proof of Lemma \ref{lem:posdefdiff}}
\begin{proof}
Since $F$ is symmetric (it is the result of subtracting two symmetric matrices), by the spectral theorem,
\begin{eqnarray}
F &=& E \Lambda E^{'} \nonumber\\
E'FE &=& \Lambda\label{eq:spectraldecomp}
\end{eqnarray}
where $\Lambda = diag(\lambda_{r},...,\lambda_{1})$ is the matrix
with eigenvalues of $F$ and $E$ is the associated eigenvector
matrix. Equality in (\ref{eq:spectraldecomp}) holds since the
eigenvalues are all distinct, implying the eigenvector matrix is
orthogonal, $E'E=I \Rightarrow E'=E^{-1}$. Note that
(\ref{eq:spectraldecomp})
implies that:
\begin{eqnarray}
\bolds{e}_{1}'F \bolds{e}_{1} &=& \lambda_{1}.\label{eq:lambda1b}
\end{eqnarray}
Assume $F \succ 0$ to prove the upper bound on $b$ and then
complete the proof by assuming the upper bound in $b$ and showing
that then $F \succ 0$. Now, given that $F$ is positive definite,
then all its
eigenvalues are positive. Therefore by (\ref{eq:lambda1b}),\\
\begin{eqnarray}
\bolds{e}_{1}^{'}F \bolds{e}_{1} &>& 0\nonumber\\
\bolds{e}_{1}'(C - bD) \bolds{e}_{1} &>& 0 \nonumber\\
b &<& \frac{\bolds{e}_{1}'C \bolds{e}_{1}}{\bolds{e}_{1}'D
\bolds{e}_{1}}.\nonumber
\end{eqnarray}
To complete the proof, assume $b <
\frac{\bolds{e}_{1}'C\bolds{e}_{1}}{\bolds{e}_{1}'D\bolds{e}_{1}}$. Then $\exists \hspace{0.1cm} a \in \Re: a > 1$ such that,\\
\begin{eqnarray}
b=\frac{v}{a(\bolds{e}_{1}'D \bolds{e}_{1})}\label{eq:expforb}
\end{eqnarray}
where $v=\bolds{e}_{1}'C \bolds{e}_{1}$. Note that $v > 0$ since $C \succ 0$. Then from (\ref{eq:lambda1b}),\\
\begin{eqnarray}
\lambda_{1} &=& \bolds{e}_{1}'(C - bD) \bolds{e}_{1}\nonumber
\end{eqnarray}
Substituting (\ref{eq:expforb}) for $b$ gives,
\begin{eqnarray}
\lambda_{1} &=&\bolds{e}_{1}'\left(C - \frac{v}{a(\bolds{e}_{1}'D
\bolds{e}_{1})}D\right) \bolds{e}_{1}\nonumber\\
&=& v - \frac{v(\bolds{e}_{1}'D \bolds{e}_{1})}{a(\bolds{e}_{1}'D
\bolds{e}_{1})} \nonumber\\
&=& (v -\frac{v}{a}) \nonumber\\
&=& v\left(\frac{(a-1)}{a}\right) > 0\nonumber
\end{eqnarray}
The last inequality is due to $v
>0$ and $a
> 1$.
\end{proof}

\subsection{Proof of Corollary \ref{cor:firstcor}}
Since, as stated in section \ref{sec:frk}, $\hat{\Sigma}_{M}$ and
$\bar{V}$ are positive definite,
$R^{-1}Q'\widehat{\Sigma}_{M}Q(R^{-1})'$ and
$R^{-1}Q'\bar{V}Q(R^{-1})'$ are also positive definite \cite[p. 141]{golubvanloan96}. The proof of the corollary is exactly
the same as for Lemma \ref{lem:posdefdiff} with $F=\hat{K},
C=R^{-1}Q'\widehat{\Sigma}_{M}Q(R^{-1})',
D=R^{-1}Q'\bar{V}Q(R^{-1})'$, and $b=\hat{\sigma}^{2}$.

\subsection{Proof of Theorem \ref{theorem:iterationalg}}
\begin{proof}
Assume $\lambda_{min,g} > \lambda_{min,g-1}$ and denote the
minimum normalized eigenvector as $\bolds{e}^{*}=\bolds{e}_{min}$, then
the eigenvector associated with $\lambda_{min,g}$ at iteration $g$
will be $\bolds{e}^{*}_{g}$. Furthermore, let $\hat{K}_{g}= C -
\hat{\sigma}^{2}_{g}D$ where we define
$C=R^{-1}Q'\widehat{\Sigma}_{M}Q(R^{-1})'$, and
 $D=R^{-1}Q'\hat{V}Q(R^{-1})'$. Showing $\hat{\sigma}^{2}_{g} <
 \hat{\sigma}^{2}_{g-1}$ is equivalent to demonstrating that if,\\
\begin{eqnarray}
  \hat{\sigma}^{2}_{g-1} =
 \hat{\sigma}^{2}_{g} + \delta_{g} \label{eq:sigmagandprev}
\end{eqnarray}
then $ \exists \hspace{0.1cm} \delta_{g} \hspace{0.1cm} \in \Re:
\delta_{g} > 0$ for all $g$. Using (\ref{eq:sigmagandprev}) one
can write,
\begin{eqnarray}
  \hat{K}_{g} &=&
 C - (\hat{\sigma}^{2}_{g-1} - \delta_{g})D\nonumber\\
 &=& \hat{K}_{g-1} + \delta_{g} D \label{eq:Khatvsprev}
\end{eqnarray}
Now one has everything required for the first part of the proof. Assuming that all eigenvalues of $\hat{K}_{g}$ are distinct, then,
\begin{eqnarray}
\bolds{e}^{*'}_{g-1}\hat{K}_{g-1}\bolds{e}^{*}_{g-1} &=&
\lambda_{min,g-1}\nonumber\\
&<& \lambda_{min,g}\nonumber\\
&=& \bolds{e}^{*'}_{g}\hat{K}_{g}\bolds{e}^{*}_{g}\nonumber\\
&\le& \bolds{e}^{*'}_{g-1}\hat{K}_{g}\bolds{e}^{*}_{g-1}\label{eq:rayleighritz}\\
&=& \bolds{e}^{*'}_{g-1}(\hat{K}_{g-1} + \delta_{g}
D)\bolds{e}^{*}_{g-1}\label{eq:rayleighritzandKhatvsprev}\\
&=& \bolds{e}^{*'}_{g-1}\hat{K}_{g-1}\bolds{e}^{*}_{g-1} +
\delta_{g} \bolds{e}^{*'}_{g-1}D\bolds{e}^{*}_{g-1}\nonumber
\label{eq:proof1a}
\end{eqnarray}
The first equality comes from (\ref{eq:lambda1b}) with
$F=\hat{K}$. The inequality afterwards is by assumption in this
part of the proof. The inequality in (\ref{eq:rayleighritz}) is
due to the Rayleigh-Ritz theorem result \citep{lutkepohl96}.
(\ref{eq:rayleighritzandKhatvsprev}) is obtained by substituting
$\hat{K}_{g}$ with (\ref{eq:Khatvsprev}). From the last equation
we see that, given that $D$ is known to be positive definite,
$\delta_{g} > 0$ for all $g$ therefore showing that
$\hat{\sigma}^{2}_{g} <
 \hat{\sigma}^{2}_{g-1}$ by (\ref{eq:sigmagandprev}).

To complete the proof, now assume $\hat{\sigma}^{2}_{g} <
\hat{\sigma}^{2}_{g-1}$
and write $\hat{\sigma}^{2}_{g} = \hat{\sigma}^{2}_{g-1} -\delta_{g}$. From (\ref{eq:Khat}) we may write,\\
\begin{eqnarray}
\hat{K}_{g} = C - \hat{\sigma}^{2}_{g}D\nonumber
\end{eqnarray}
Then,\\
\begin{eqnarray}
\hat{K}_{g} &=& C - (\hat{\sigma}^{2}_{g-1} -\delta_{g})D\nonumber\\
 &=& C - \hat{\sigma}^{2}_{u,g-1}D +\delta_{g}D\nonumber\\
&=& \hat{K}_{g-1} + \delta_{g}D\nonumber
\end{eqnarray}
from this result we deduce that
$\lambda_{min}(\hat{K}_{g})=\lambda_{min}(\hat{K}_{g-1} +
\delta_{g}D)$. As a final step,\\
\begin{eqnarray}
\lambda_{min}(\hat{K}_{g}) &=& \lambda_{min}(\hat{K}_{g-1} +
\delta_{g}D)\nonumber\\
&>& \lambda_{min}(\hat{K}_{g-1})\nonumber
\end{eqnarray}
the last inequality is due to theorem 8.1.5 in \cite[p. 396]{golubvanloan96}.
\end{proof}

\subsection{Proof of Corollary \ref{cor:othercor} and Corollary \ref{theorem:iterationalgweighted}}
Proofs are almost identical to Lemma \ref{lem:posdefdiff} and
Theorem \ref{theorem:iterationalg} respectively.

\bibliographystyle{ieeetran}
\bibliography{bibtexreferences2011}

\end{document}